\documentclass[prd,preprint,tightenlines,floatfix,showpacs,preprintnumbers,nofootinbib,eqsecnum]{revtex4}

\pdfoutput=1

 \usepackage[dvips,final]{graphicx}
  \usepackage{amssymb}
   \usepackage{amsmath}
    \usepackage{amsfonts}
     \usepackage{epsfig}
      \usepackage{bm} 
      \usepackage[dvipsnames,usenames]{color}
      \usepackage{comment}
      \usepackage{float}
      
      \usepackage[utf8]{inputenc}

\begin{document}

\title{DPS mechanism for associated $c\bar c l^+l^-$ production 
in $AA$ UPCs as a probe for photon density inside the nucleus}

\author{Edgar Huayra$^{1}$}
\email{yuberth022@gmail.com}

\author{Emmanuel G. de Oliveira$^{1}$}
\email{emmanuel.de.oliveira@ufsc.br}

\author{Roman Pasechnik$^{1,2}$}
\email{Roman.Pasechnik@thep.lu.se}

 \author{Bruna O. Stahlh\"ofer$^{1}$}
\email{stahlhoferbruna@gmail.com}

\affiliation{
\\
{$^1$\sl Departamento de F\'isica, CFM, Universidade Federal 
de Santa Catarina, C.P. 476, CEP 88.040-900, Florian\'opolis, 
SC, Brazil
}\\
{$^2$\sl
Department of Astronomy and Theoretical Physics, Lund
University, SE-223 62 Lund, Sweden
}}

\begin{abstract}
\vspace{0.5cm}
We discuss the associated $c\bar{c}$ and $l^+l^-$ pairs production in ultraperipheral heavy-ion collisions at high energies. Such a channel
provides a novel probe for double-parton scattering (DPS) at small $x$ enabling one to probe the photon density inside the nucleus. We have derived an analog of the standard central $pp$ pocket formula and studied the kinematical dependence of the effective cross section. Taking into account both elastic and non-elastic contributions, we have shown predictions for the DPS $c\bar c l^+l^-$ production cross section differential in charm quark rapidity and dilepton invariant mass and rapidity for LHC and a future collider.
\end{abstract}


\maketitle

\section{Introduction}
\label{Sect:intro}

Particle production processes driven by multi-parton interactions in the high energy proton-proton, proton-nucleus or nucleus-nucleus collisions provide an opportunity to probe the inner structure of colliding protons and nuclei. In the double-parton scattering (DPS), for instance, one considers two partons (typically, quarks or gluons) of a colliding nucleus (or proton) interacting independently with the other particle in the same collision. A comprehensive review can be found in Ref.~\cite{Manohar:2012pe, Diehl:2011yj, Bartalini:2011jp, Diehl:2017kgu, Bansal:2014paa} and recent experimental results, e. g., in Refs.~\cite{Aaij:2020smi, Aaboud:2018tiq, Sirunyan:2017hlu}. Double parton scattering is specially sensitive to the correlations in the impact parameter distribution of the two partons. As an effect, most of the DPSs are similar in the sense that quarks and gluons roughly share the same impact parameter distribution that shows little to no dependence on longitudinal momentum fraction $x$.

In the previous work of Ref.~\cite{Huayra:2019iun} by some of the authors, the associated production of $c\bar c$ and $b\bar b$ pairs well-separated in rapidity has been studied in high-energy $pA$ ultra-peripheral collisions (UPCs). This observable mainly probes the Weiszäcker-Williams (WW) ~\cite{vonWeizsacker:1934nji, Williams:1934ad, Williams:1935dka} double-photon distribution in the projectile nucleus and double-gluon parton distribution in the target proton. By taking the well understood WW photons that are more broadly distributed in impact parameter than the usual quarks and gluons, this process becomes very different from the standard (or central collision) DPS and allows for an independent way of exploring the parton correlations in the target. In a follow up work of Ref.~\cite{Huayra:2020iib}, the same reaction has been employed to explore the double-gluon distribution in the nucleus considering $AA$ UPCs, when the target nucleus was allowed to dissociate. In both cases, the resulting cross sections were large enough to have produced a few events at current accelerators and one hopes that, in the future, controlling for the background and estimating theoretical errors, new information about the correlations involving partons will be available with these observables. Also, recently, a new study of a $\gamma p$ double parton cross section was proposed in Ref.~\cite{Rinaldi:2021vbj}.
\begin{figure}[tb]
	\centering
	\includegraphics[trim= 0cm 5cm 0cm 5cm,clip, width=.6 \textwidth]{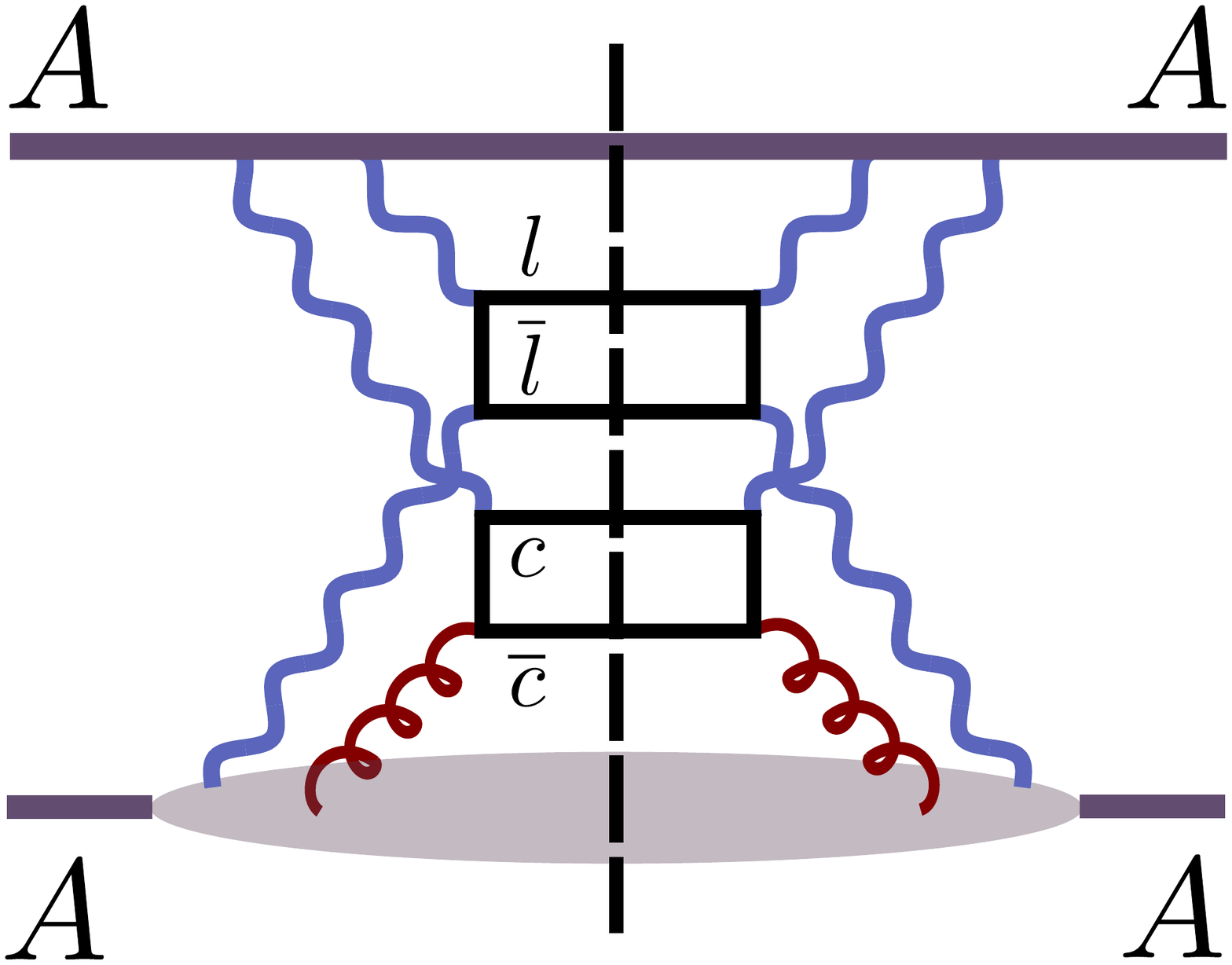}
	\caption{An illustration of the cross section in the DPS mechanism of the dilepton $l^+l^-$ and $c\overline{c}$ quark pairs production through the dominant mechanism involving three photons and one gluon in the initial state. For the upper nucleus, we use punctual charge photon distribution while for the lower one we apply the Woods-Saxon charge distribution (for the elastic component) and the photon PDF (for the inelastic component).}
	\label{fig:3f1g}
\end{figure}

In the present work, we continue our investigation of the internal nuclear structure through DPS in $AA$ UPCs and consider a mixed lepton-quark production mode. More specifically, we consider the associated production of the heavy-quark pair, e.g. $c\bar c$, and a leptonic pair $l^+l^-$. At the DPS parton level, this reaction is prompted by two hard subprocesses ($\gamma\gamma \to l^+l^-$ and $\gamma g \to c\bar c$) occurring in the same $AA$ collision, i.e., this production will be dominated by an initial state with three photons and one gluon (see such an illustration in Fig.~\ref{fig:3f1g}). Therefore, we aim at studying the photon--gluon distributions in the target by using the WW distribution of the projectile that will not be allowed to dissociate. At the best of our knowledge, this reaction has not been studied in the context of double parton scattering. We estimate both the elastic and inelastic contributions to the DPS cross section and conclude that the magnitude of photon--gluon correlation effects is relatively small and does not exceed 10\%. So, this justifies the basic approximation adopted in our analysis neglecting the initial-state correlations in practical calculations.

It is important to realize that, since a gluon is taken from the target, it will most likely dissociate. Therefore, we can consider two kinds of photon distributions from the target: usual WW used in UPCs but also the photon as a parton with a transverse coordinate distribution similar to the other partons. In the case of a nucleus as the target, the first one will dominate and than we will have a DPS not like the usual 4--gluon case and also dissimilar to the 2-photon--2-gluon case introduced in Ref.~\cite{Huayra:2019iun}. Nevertheless, WW photons in UPCs are usually taken to be outside the nucleus, while in this work the photon distribution will be probed in the target also inside of the nucleus. The latter is yet poorly known and is a subject of intense discussions in the literature, specially in the context of dilepton production, see e.g. \cite{KlusekGawenda:2010kx,Klusek-Gawenda:2020eja,Harland-Lang:2021ysd,Brandenburg:2021lnj}.

This article is organised as follows. In Section~\ref{Sect:inside}, we discuss the modelling of photon density inside the target nucleus. In Section~\ref{Sect:dilepton}, we provide an overview of the single parton scattering (SPS) process for dilepton pair production in $\gamma\gamma$-fusion in $AA$ UPCs when the projectile nucleus survives intact the collision while the target nucleus can dissociate. In Section~\ref{Sect:charm} we do likewise for the $c\overline{c}$ production. In Section~\ref{Sect:predictions}, we evaluate the DPS effective cross section of the associated $c\bar{c}$ and $l^+l^-$ pairs production and also provide various predictions for such observable in AA and Ap collisions. A summary and concluding remarks are given in Section~\ref{Sect:concl}.

\section{Photon flux of the nucleus}
\label{Sect:inside}

In calculations of $AA$ ultraperipheral collisions (UPCs), the impact parameter is usually required to be larger than the sum of the two nuclear radii, $b > 2R_A$. In that sense, this kind of collision is governed by electromagnetic interactions. Such interactions happen due to the cloud of virtual photons produced by the charges of the ultra-relativistic moving nucleus~\cite{Vogt:2007zz}. The photon flux induced by the charges is estimated using the Weizs\"acker-Williams method \cite{vonWeizsacker:1934nji, Williams:1934ad, Williams:1935dka,Jackson:1998nia} and, in most cases, will leave the nucleus intact, therefore they are called ``elastic'' photons.

For a point charge, the standard WW photon flux as a function of photon energy $\omega$ and impact parameter $\vec{b}$ is written as 
\begin{align}
\label{photonDist_b}
     \frac{d^3 N (\omega, \vec{b} )}{d \omega d^2 \vec{b} } 
    = \frac{ \alpha Z^2 k^2}{\pi^2 \omega b^2 } \left[ K_1^2 (k) 
    + \frac{1}{\gamma^2} K_0^2 (k) \right] \,, \quad k = \frac{b \omega}{\gamma} \,,
\end{align}
and the corresponding $b$-integrated flux reads
\begin{align}
\label{photonDist}
     \frac{d N (\omega)}{d \omega} = \frac{2 \alpha Z^2}{ \pi \omega } \left[ \zeta K_0 (\zeta) K_1 (\zeta)
    - \frac{\zeta^2}{2} \left( K_1^2 (\zeta) - K_0^2 (\zeta) \right) \right] \,,
\end{align}
where $\zeta = \omega b_{\text{min}}/\gamma$, $\alpha=1/137$ is the fine structure constant and $K_{0,1}$ are modified Bessel functions. The lower limit of $b$ integration is the radius of the lead nucleus $ b_{\text{min}} \equiv R_{\text{Pb}} = 6.63$ fm. Here, using the proton 
mass $m_p = 0.938$ GeV, we can set the Lorentz factor to $\gamma = \sqrt{s}/2  m_p$, where $\sqrt{s}$ is the center-of-mass (c.m.) energy of the nucleon--nucleon collision. For a typical LHC energy of $\sqrt{s} = 5.02$ TeV, we have $\gamma \approx 2676$. If the photon distribution is probed far away from the nucleus, the point charge approximation is good enough.

\begin{figure}[tb]
\centering
\includegraphics[width=.6\textwidth]{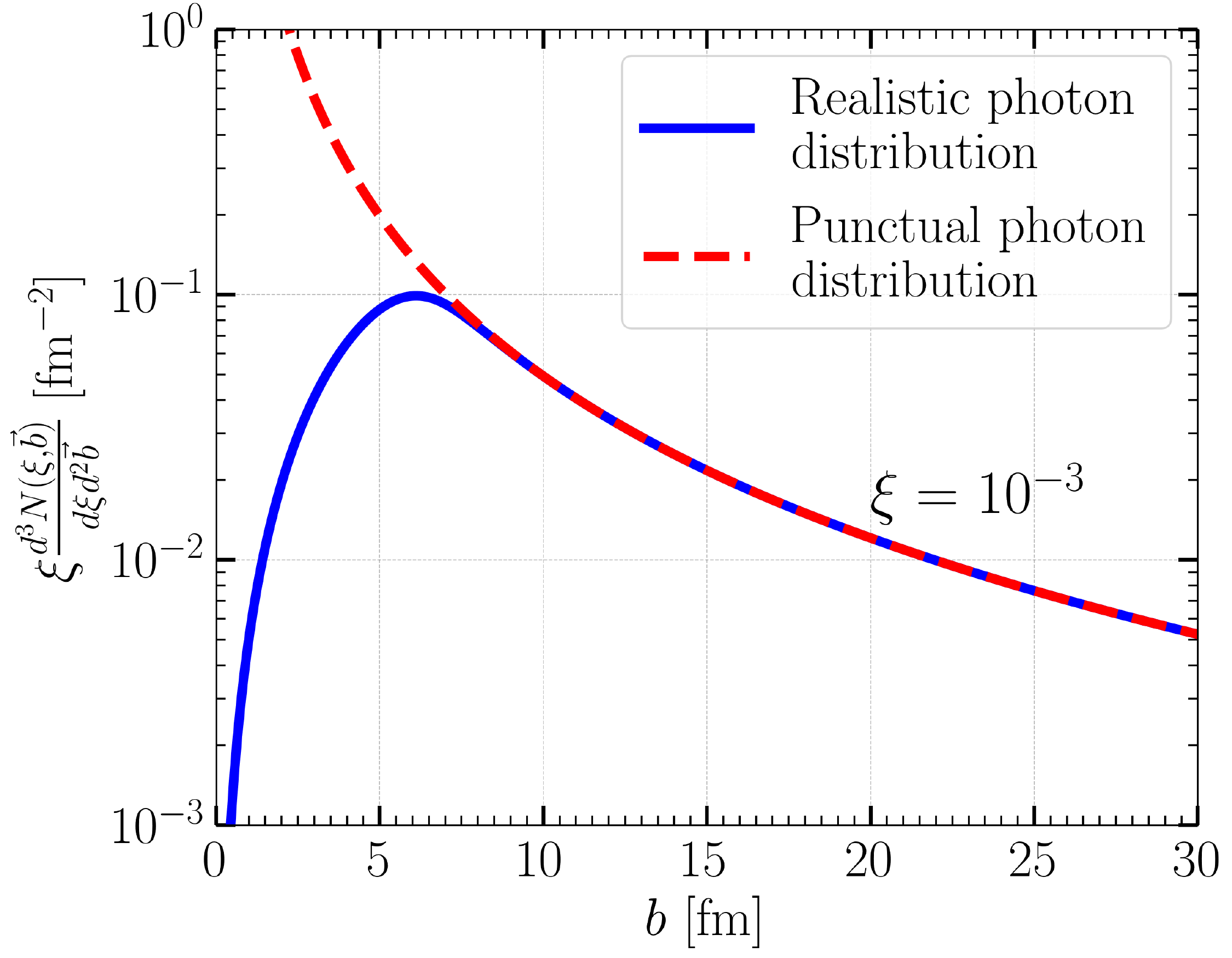}
\caption{\label{fig:DistFotons}
The photon number distribution as a function of impact parameter $b$ and with energy fraction fixed to $\xi=10^{-3}$ showing the difference between point like and realistic charge distributions.}
\end{figure}

However, the nucleus is a complex object and there will be cases that we will be interested in the photon distribution very close to or even inside the nucleus. Therefore, it is better to consider a realistic charge distribution of the nucleus to obtain the photon flux. In order to do so, we must remember that the photon flux as a function of impact-parameter and of photon energy is obtained from
\begin{align}
	\omega \, \frac{d^3N (\omega,{\vec b})}{d\omega d^2{\vec b}}  = \frac{Z^2 \alpha}{\pi^2} \left| \int_0^{\infty} 
	d q_t \frac{q_t^2 \text{F} \left( q_t^2 + \frac{\omega^2}{\gamma^2} \right) }
	{ q_t^2 + \frac{\omega^2}{\gamma^2} } \text{J}_1 (b q_t) \right|^2 \,,
\label{eq:photondistribution}
\end{align}
where $\mathbf{q}$ is the photon momentum, $q_t^2 + \frac{\omega^2}{\gamma^2} = \mathbf{q}^2$, $\text{J}_1$ is the Bessel function of the first kind. The form factor of the nucleus $F(\mathbf{q}^2)$ is found in terms the  by applying the Fourier transform to the spherically symmetric charge distribution $\rho(r)$:
\begin{align}
F(\mathbf{q}^2) = 4 \pi \int_0^{\infty} r d r \, \rho(r) \,
\frac{ \sin(|\mathbf{q}| r)} {|\mathbf{q}|}.
\label{eq:Fourier transform}
\end{align}

In this work, we explore observables in lead-lead (PbPb) UPCs (with atomic weight $A=208$ and number $Z=82$), 
and we adopt the following normalised Wood-Saxon parameterization~\cite{Bromley:1967ixa}
\begin{align}
\rho(\vec r) = \frac{\rho_0}{1 + \exp\left( \frac{r - R_A}{\delta} \right) } \,, \qquad 
\int d^3 r \, \rho(\vec r) = 1 \,,
\label{eq:WoodSaxon}
\end{align} 
where $\delta = 0.459$ fm and the nuclear radius is $R_A = 6.63$ fm, and $\rho_0$ is an overall normalization. More details about parameterizations of the charge distribution in the nucleus can be found in, e.g., Ref.~\cite{KlusekGawenda:2010kx, Mariola:Thesis}).

It is convenient to express the photon flux as a distribution in terms of the energy fraction carried by the photon $\xi$ of the nucleus
\begin{eqnarray}
\xi \, N (\xi,{\vec b})
\equiv
\xi \, \frac{d^3 N(\xi,{\vec b})}{d\xi d^2{\vec b}}= 
\omega \, \frac{d^3N (\omega,{\vec b})}{d\omega d^2{\vec b}} 
\qquad \text{with} \qquad 
\xi = \frac{2 \omega}{\sqrt{s}} \,.
\end{eqnarray}
Later, it will be useful to integrate over the impact parameter $\vec b$ as follows,
\begin{align}
\xi \bar{N}_{\gamma} (\xi) \equiv \xi \frac{d N}{d \xi}
= \int d^2 b \, \xi \, \frac{d^3 N(\xi,{\vec b})}{d\xi d^2{\vec b}}. 
\label{eq:Nbar}   
\end{align}

This Wood-Saxon parameterization provides a realistic photon distribution as illustrated in Fig.~\ref{fig:DistFotons} (solid line). We see a maximum of the distribution at approximately $6.63$ fm, i.e., close to the radius of the nucleus. For very small values of the impact parameter as we move towards the nuclear center, the number of photons drops to zero as expected, unlike the distribution by a point charge (dashed line) in which we notice the typical divergence at the origin.

\begin{figure}[tb]
\centering
\includegraphics[width=.6\textwidth]{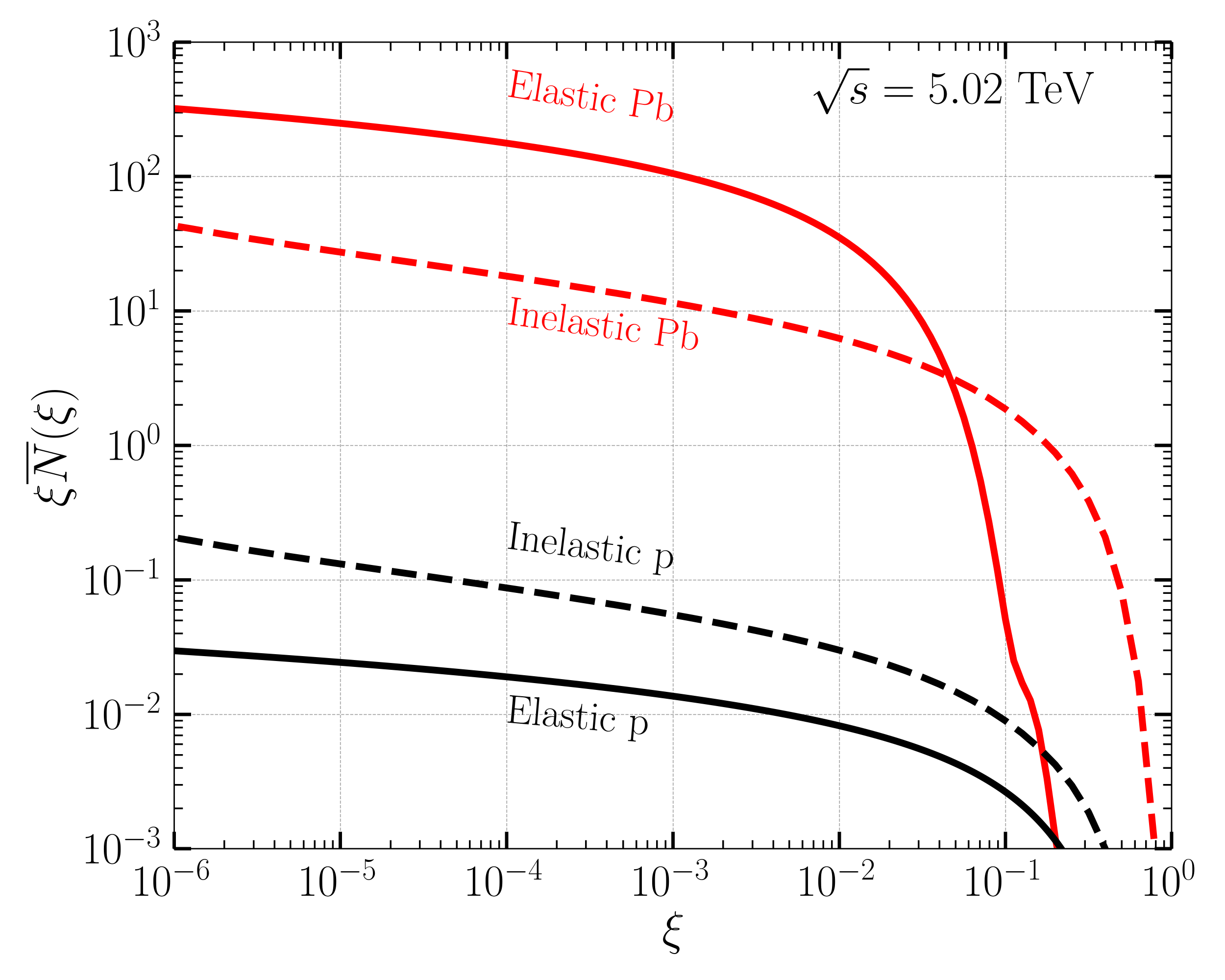}
\caption{The elastic and inelastic photon distributions integrated over impact parameter $b$ as a function of energy fraction $\xi$ carried by the photon of the proton and the nucleus.}
\label{fig:N_barXi_real}
\end{figure}

To have a more complete treatment of the UPCs $AA$, besides the elastic contribution to the photon distribution, we also consider the inelastic one arising due to the partonic evolution in the nucleus, such photons are treated as nucleon constituents. We will use the photon PDF parameterization ``MMHT2015qed''~\cite{Harland-Lang:2019pla} for the proton an we will multiply it by $A$ for the nucleus. We are not including nuclear modification factors to the inelastic photon density since we expect less photon shadowing effect (and other effects) than in the case of the gluon distribution. Besides, the inelastic contribution is relatively very small and will not have a noticeable impact on our calculated observables when the target is the nucleus. For a thorough discussion of prospects for measurements of these distributions in collider experiments such as ATLAS, see Ref.~\cite{Trzebinski:2019tmk}.

In order to study the proton as a target in pA ultraperipheral collisions, we need to estimate the elastic contribution to the photon flux of the proton. In the case for a point-like charge parameterization of the proton, we use the equation \eqref{photonDist} with $Z = 1$ and with the minimum value of the impact parameter taken to be the proton radius $b_{\text{min}} \equiv R_p = 0.84$ fm \cite{Guzey:2014axa}. For a more realistic elastic photon distribution of the proton, we consider the following model~\cite{Drees:1988pp, daSilveira:2021bzs}: 
\begin{align}
    \overline{N}_{\gamma|p} (\xi) 
    = \frac{\alpha}{2 \pi \xi} [1 + (1 - \xi)^2] 
    \left( \ln \Omega - \frac{11}{6} + \frac{3}{\Omega} - \frac{3}{2 \Omega^2} + \frac{1}{3 \Omega^3} \right) \,,
\end{align}
where $\Omega = 1 + (0.71 \text{GeV}^2) / Q^2_{\text{min}}$, and 
\begin{align}
    Q^2_{\text{min}} \approx 
    \frac{(\xi m_p )^2}{1 - \xi} \,.
\end{align}
Recently, the work of Ref.~\cite{daSilveira:2021bzs} explored both the elastic and inelastic contributions for $pp$ collisions. 

The two contributions (elastic and inelastic) to the photon flux of
the proton and the nucleus are illustrated in the Fig.~\ref{fig:N_barXi_real}. We can see that the elastic photon distribution is greater than the inelastic photon distribution in the case of the nucleus for small values of $\xi$, unlike the proton, for which inelastic distribution is greater than the elastic distribution.

\section{Dilepton production in UPCs}
\label{Sect:dilepton}

In order to compute the SPS differential cross section of the dilepton production in $AA$ UPCs, 
we start with the partonic $\gamma_1 \gamma_2 \rightarrow l\bar{l}$ 
cross section (Ref.~\cite{Vogt:2007zz} is one review among others):
\begin{align}
\frac{d^2 \hat{\sigma}_{\gamma \gamma \rightarrow l\bar{l}}}{d\hat{t} d\hat{u}} = \frac{2 \pi \alpha^2}{\hat{s}^2} \left( \frac{\hat{t}}{\hat{u}} + \frac{\hat{u}}{\hat{t}} + \frac{4 m_l^2 \hat{s}}{\hat{t} \hat{u}} 
\left( 1 - \frac{m_l^2 \hat{s}}{\hat{t} \hat{u} } \right) \right) 
\delta (\hat{s} + \hat{t} + \hat{u}) \,,
\end{align}
where we use the modified Mandelstam variables, $ \hat{t} = (p_l - p_{\gamma_1})^2 - m_l^2 $, 
$ \hat{u} = (p_{\bar l} - p_{\gamma_1})^2 - m_l^2 $ and 
$\hat{s} = (p_l + p_{\bar l})^2 = M^2$. Here, $m_l$ and $M$ are the lepton 
and dilepton masses, respectively. The corresponding cross section differential 
in the lepton transverse momentum reads
\begin{align}
\frac{d \hat{\sigma}_{\gamma \gamma \rightarrow l\bar{l}}}{d p_{l\perp}^2}
= \frac{1}{\sqrt{1 - 4 (m_l^2 + p_{l\perp}^2)/M^2}} 
\frac{d \hat{\sigma}_{\gamma \gamma \rightarrow l\bar{l}}}{d\hat{t}} \,,
\end{align}
that can be utilised to compute the parton-level cross section differential in dilepton
rapidity ($Y$) and invariant mass ($M^2$) as follows
\begin{eqnarray}
\frac{d^2\hat{\sigma}_{\gamma \gamma \rightarrow l\bar{l}}}{dYdM^2} 
= \frac{\xi_1 \xi_2}{M^2} \int d p_{l\perp}^2 \frac{d\hat{\sigma}_{\gamma \gamma \rightarrow l\bar{l}}}{d p_{l\perp}^2} \delta\left( \xi_1 - \frac{M}{\sqrt{s}} \mathrm{e}^{Y} \right) 
\delta\left( \xi_2 - \frac{M}{\sqrt{s}} \mathrm{e}^{-Y} \right) \,.
\end{eqnarray}

\begin{figure}[tb]
\centering
\includegraphics[width=.48\textwidth]{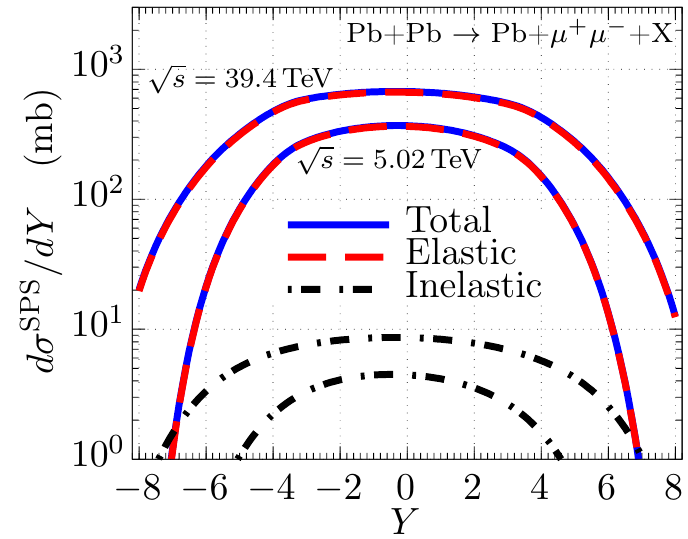} \hfill
\includegraphics[width=.48\textwidth]{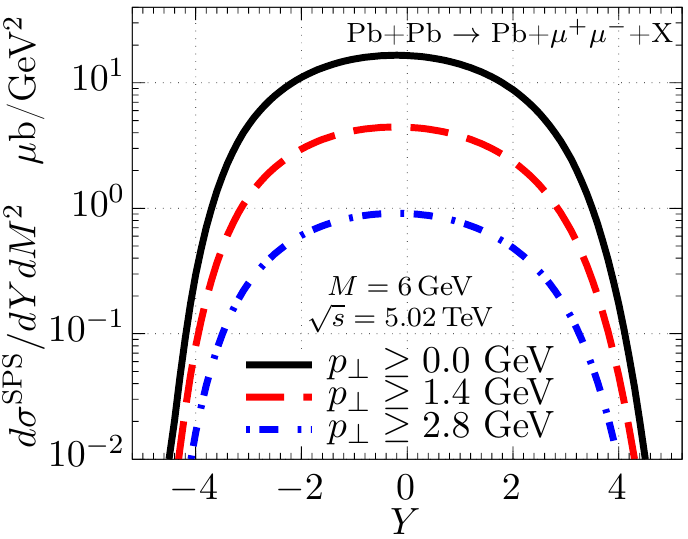}
\caption{The SPS dimuon production cross section in PbPb UPCs at typical LHC ($\sqrt{s}=5.02$ TeV) and FCC ($\sqrt{s}=39.4$ TeV) energies as a function of dimuon rapidity $Y$ and integrated over dimuon invariant mass $M$ (left panel) or at a fixed $M = 6$ GeV with a muon transverse momenta lower cutoff (right panel).}
\label{fig:SPSdilepton}
\end{figure}

Turning to the hadron level, the differential dilepton cross section in $AA$ UPCs is 
given by
\begin{eqnarray}
\nonumber
\frac{d^2\sigma_{AA \rightarrow AX + l\bar{l}}}{dY dM^2} 
& = & \int d^2 \vec{b} d^2 \vec{b}_{\gamma_1} d^2 \vec{b}_{\gamma_2} d \xi_1 d \xi_2\, 
\Theta(b - 2R_A) 
\delta^{(2)} (\vec{b} + \vec{b}_{\gamma_2} - \vec{b}_{\gamma_1}) \\
& \times & 
\, \Theta(b_{\gamma_1} - R_A) N_{\gamma_1}(\xi_1,\vec{b}_{\gamma_1}) N_{\gamma_2}(\xi_2,\vec{b}_{\gamma_2}) 
\frac{d^2\hat{\sigma}_{\gamma_1 \gamma_2 \rightarrow l\bar{l}}}{dYdM^2} \, .
\label{eq:SPSdileptonfirst}
\end{eqnarray}
where $N_{\gamma_1}(\xi_1,\vec{b}_{\gamma_1})$ and $N_{\gamma_2}(\xi_2,\vec{b}_{\gamma_2})$
are the photon distributions in the upper and lower nucleus lines
illustrated in Fig.~\ref{fig:3f1g}. The main difference between these is that the punctual charge
photon distribution is used in the upper nucleus $N_{\gamma_1}$, while the Woods-Saxon charge
distribution (elastic) and the photon PDF in the nucleon (inelastic) are applied as 
the photon density $N_{\gamma_2}$ in the lower nucleus. 
In Eq.~(\ref{eq:SPSdileptonfirst}), one performs an integration over the photon momentum 
fractions $\xi_{1,2}$ and also the impact parameters -- the relative one $\vec{b}$ 
and the impact parameters of the photons $\vec{b}_{\gamma_{1,2}}$, using the Heaviside function $\Theta(b-2R_A)$ 
in order to ensure an ultraperipheral nature of $AA$ collisions. In Fig.~\ref{fig:SPSdilepton}, we illustrate some of the results that can be obtained with the above formula, in order to have a basis of comparison when discussing the DPS case.

\section{$c\bar{c}$ production in UPCs}
\label{Sect:charm}

\begin{figure}[tb]
\centering
\includegraphics[width=.6\textwidth]{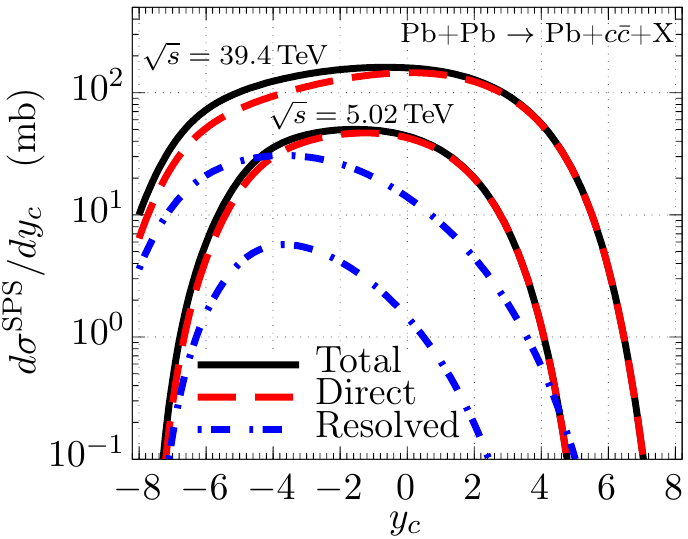}
\caption{The SPS $c \bar c$ production cross section in PbPb UPCs at typical LHC ($\sqrt{s}=5.02$ TeV) and FCC ($\sqrt{s}=39.4$ TeV) energies integrated in all $p_{c\perp}$ as a function of rapidity $y_c$.}
\label{fig:SPScharm}
\end{figure}

The $AA$ UPC SPS cross section for $c\bar{c}$ pair production differential in $c$-quark transverse momentum and $c$- and $\bar c$-quark rapidities can be written as follows
\begin{align} \label{ccbar-AA-CS} \nonumber
\frac{d^3\sigma_{AA \rightarrow AX + c\bar{c}}}{dy_{c}dy_{\bar{c}}dp_{c\perp}^2} 
 & = 
\int d^2 \vec{b} d^2 \vec{b}_\gamma  d^2 \vec{b}_g d \xi d x\, 
\Theta(b - 2R_A) \delta^{(2)} (\vec{b} + \vec{b}_g - \vec{b}_\gamma) \\ & \times  
\Theta(b_{\gamma} - R_A) N_{\gamma_3}(\xi,\vec{b}_\gamma) G^{\rm A}_g(x, \vec{b}_g) 
\frac{d^3\hat{\sigma}_{\gamma g \rightarrow c\bar{c}}}{dy_{c}dy_{\bar{c}}d p_{c\perp}^2} \,.
\end{align}
In the above formula we have the convolution of parton distributions with the parton level cross section with the UPC requirement given by the Heaviside function $\Theta(b - 2R_A)$. Some numerical results are shown in Fig.~\ref{fig:SPScharm}. 

The differential parton-level $\gamma+g \to c\bar{c}$ cross section can be written as 
\begin{align}
\frac{d^3\hat{\sigma}_{\gamma g \rightarrow c\bar{c}}}{dy_{c}dy_{\bar{c}}d p_{c\perp}^2}
& = \frac{1}{\sqrt{1- 4(m^2_c+p^2_{c\perp})/\hat{s}}}  \frac{d\hat{\sigma}_{\gamma g \rightarrow c\bar{c}}}{d\hat{t}}
 \, \delta\left (y_{c} - \frac12 \ln \left( \frac{\xi\hat{u}}{x\hat{t}} \right) \right) 
 \, \delta\left (y_{\bar{c}} - \frac12 \ln \left( \frac{\xi\hat{t}}{x\hat{u}} \right) 
 \right) \,,
 \end{align}
where we have used again the modified Mandelstam variables, e.g., $\hat{t} \equiv (p_c - p_\gamma)^2 - m^2_c$, and
\begin{eqnarray}
\frac{d^2\hat{\sigma}_{\gamma g \rightarrow c\bar{c}}}{d\hat{t}d\hat{u}} = 
\frac{\pi \alpha_s \alpha e^2_c}{\hat{s}^2} \bigg[ \frac{\hat{t}}{\hat{u}}+
\frac{\hat{u}}{\hat{t}}+\frac{4m^2_c\hat{s}}{\hat{t}\hat{u}}
\bigg(1-\frac{m^2_c\hat{s}}{\hat{t}\hat{u}}\bigg) \bigg] 
\delta(\hat{s}+\hat{t}+\hat{u}) \,.
\end{eqnarray}
We set the charm quark mass to $m_c=1.4$ GeV and the charge to $e_c = 2/3$.

We will use again the the photon distribution $N_\gamma(\xi,\vec{b}_\gamma)$ from a point charge on the projectile side, with longitudinal momentum fraction $\xi$ and impact parameter $\vec{b}_\gamma$. On the target side, we have the gluon distribution of a nucleus, with longitudinal momentum fraction $x$ and impact parameter $\vec{b}_g$, that is built from the nucleon distribution by factorizing the $x$ and  $\vec{b}_g$ dependencies:
\begin{align} 
 G^{\rm A}_g(x, \vec{b}_g)
 & = 
A g(x) \int d^2 \vec{b}_p \delta^{(2)} (\vec{b}_g - \vec{b}_p - \vec{b}_{g|p})  \rho_{\rm 2D}(\vec b_p)
 f_g(\vec{b}_{g|p}).
\end{align}
For the collinear gluon density $g(x)$ of the nucleon in a given nucleus $A$ we use the EPPS16nlo parameterisation \cite{Eskola:2016oht}, with the factorisation scale being the hard scale of the subprocess $\mu_F^2 \equiv \hat{s}$. 

To localize the nucleon inside the nucleus in impact parameter space, the two-dimensional thickness function is given by the integral over the longitudinal coordinate $z$ of the same Wood--Saxon distribution used before:
\begin{align}
\rho_{\rm 2D}(\vec b) \equiv \int dz\, \rho(\vec r) = \int dz\, \frac{\rho_0}{1+\exp\Big((\sqrt{b^2 + z^2} - R_A)/\delta\Big)} \,,
\label{eq:WoodSaxon-2D}
\end{align}
where $\vec{b}_p$ is the position of such nucleon. To get the gluon impact parameter distribution in a nucleus, the above is convoluted with the profile function~\cite{Frankfurt:2010ea}, 
\begin{align}
    f_g (\vec{b}) =
    \frac{\Lambda^2}{2 \pi} \frac{\Lambda b}{2} K_1 (\Lambda b), 
    \qquad 
    \int d^2 \vec{b} f_g (\vec{b}) = 1 \,,
\end{align}
being the normalised spatial gluon density of a nucleon as explained in Ref.~\cite{Huayra:2019iun}. Here, $\Lambda \approx 1.5$ GeV is the scale parameter of the distribution, and $K_1$ is the modified Bessel function.

\section{DPS mechanism and numerical results}
\label{Sect:predictions}

\begin{figure}[tb]
\centering
\includegraphics[width=.75\textwidth]{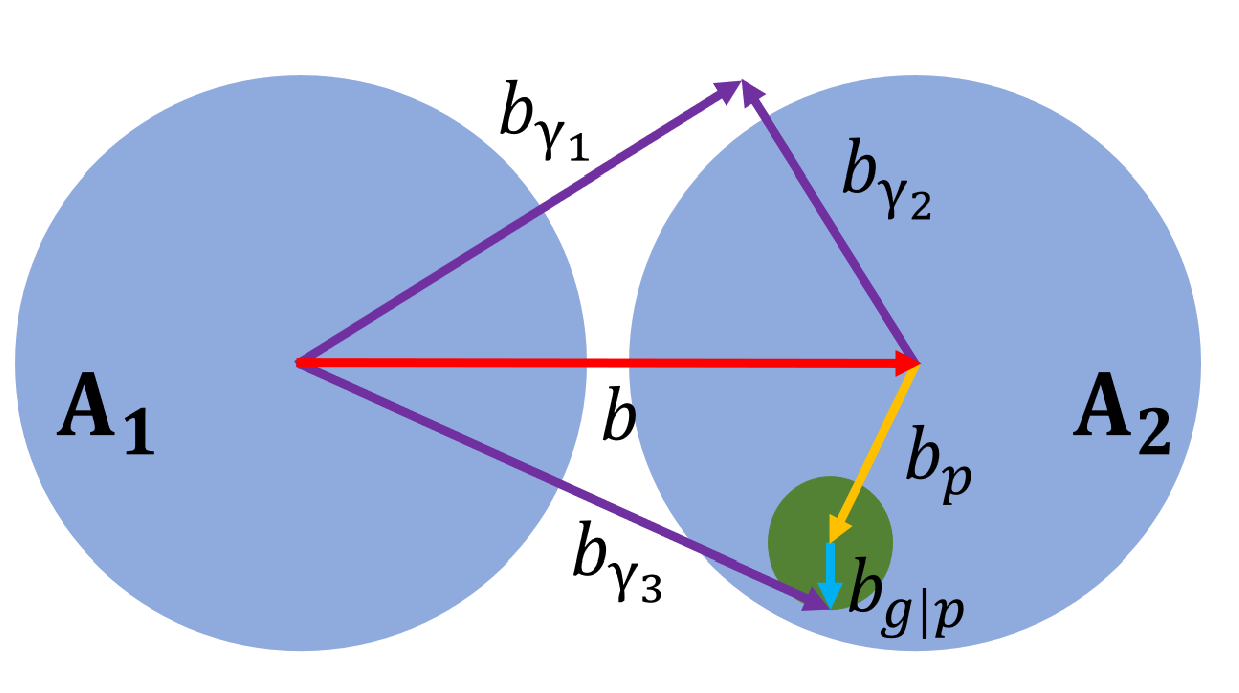}
\caption{The impact parameter picture of DPS in $AA$ UPCs. Two photons from the projectile interact with one gluon and one photon from the target.}
\label{fig:fig_AA_3f1g}
\end{figure}

\begin{figure}[tb]
\centering
\includegraphics[width=.6\textwidth]{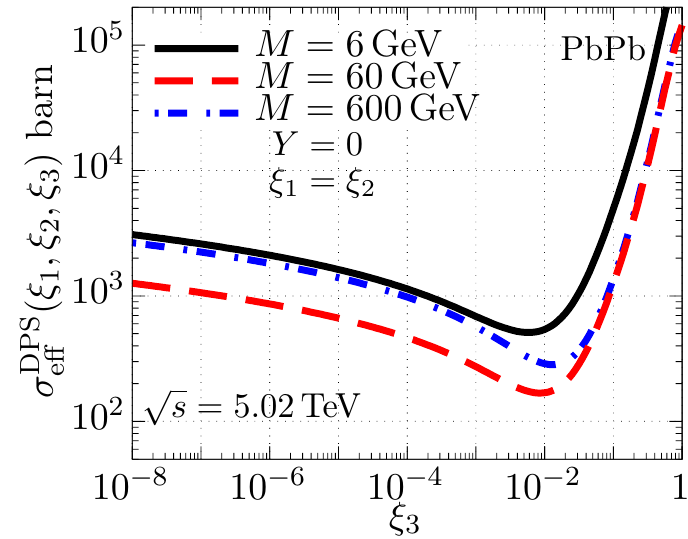}
\caption{The DPS effective cross section $\sigma^{\rm PbPb}_{\rm eff}$ for $3\gamma + g$ 
in the initial state with $M = 6$, $M=60$ and $M=600$ GeV and $Y = 0$.}
\label{fig:sigeff}
\end{figure}

The general formula for the double parton scattering involving the $\gamma g \rightarrow c \bar c$ and $\gamma \gamma \rightarrow l \bar l$ subprocesses in a UPS collision in which one of the nucleus only provides photons and likely survives the interaction (projectile) while the other likely breaks up after providing a gluon and a photon (target) is the following:
\begin{align}
\nonumber
\frac{d^5 \sigma^\text{DPS}_{AA\rightarrow AXc\bar{c}l\bar{l}}}{d y_c 
d y_{\bar{c}} d p_{c\perp}^2 \,d Y d M }
& = \int d^2 \vec{b} \, \Theta(b - 2R_A) \int d^2 \vec{b}_{\gamma_1} 
\int d^2 \vec{b}_{\gamma_3} \int d \xi_1 d \xi_2 d \xi_3 d x  \\
& \times N_{\gamma_1 \gamma_3} (\xi_1, \vec{b}_{\gamma_1}; \xi_3, \vec{b}_{\gamma_3}) 
P^{\rm A}_{\gamma_2 g} (\xi_2, \vec{b}_{\gamma_2}; x, \vec{b}_g) 
 \frac{d \hat{\sigma}_{\gamma g \rightarrow c\bar{c}}}{d p_{c\perp}^2}
\int \frac{d p_{l\perp}^2}{M^2} \frac{d \hat{\sigma}_{\gamma \gamma \rightarrow l\bar{l}}}{d p_{l\perp}^2}\,.
\end{align}
In the limit of high energies, it is safe to neglect the correlations between the two projectile photons:
\begin{eqnarray} 
    N_{\gamma_1\gamma_3} (\xi_1, \vec{b}_{\gamma_1}; \xi_2, \vec{b}_{\gamma_3}) & = & 
    \Theta (b_{\gamma_1} - R_A)  \Theta (b_{\gamma_3} - R_A) N_{\gamma_1} (\xi_1, \vec{b}_{\gamma_1}) \,
    N_{\gamma_3} (\xi_3, \vec{b}_{\gamma_3}) 
    \label{Nfactorisation}
\end{eqnarray}

On the other hand, correlations between gluons and photons in the target are more interesting. For the sake simplicity, we also assume factorization in what follow:
\begin{eqnarray} 
    P^{\rm A}_{\gamma_2 g} (\xi_2, \vec{b}_{\gamma_2}; x, \vec{b}_g) & = & 
    N_{\gamma_2} (\xi_2, \vec{b}_{\gamma_2}) \,
    G^{\rm A}_{g} (x, \vec{b}_g)
    \label{Pfactorisation}
\end{eqnarray}
Nevertheless, we will also consider the inelastic contribution to the DPS. In this, the photon is as much of a parton as the gluon and its impact parameter distribution is the same. In this case, we consider that inside the nucleon, the two partons are uncorrelated. However, we consider the correlation created by  the two partons coming from the same nucleon, as opposed to the case where they come from different nucleons. We call this the inelastic contribution as as we will see, it is about 10\% or less than the elastic one. The precise calculation of these contributions are already established in our previous paper \cite{Huayra:2019iun}.

In analogy to the standard pocket formula, the differential DPS cross section 
can be represented as a product 
of auxiliary cross sections of $\gamma g \rightarrow c\bar{c}$ and 
$\gamma \gamma \rightarrow l\bar{l}$ subprocesses as follows
\begin{align}
\frac{d^5 \sigma^\text{DPS}_{AA\rightarrow AXc\bar{c}l\bar{l}}}{d y_c 
d y_{\bar{c}} d p_{c\perp}^2 \,d Y d M }
= \frac{ 1 }{\sigma^{AA}_\text{eff}(\xi_1, \xi_2, \xi_3)}
\frac{d^3 \Sigma_{\gamma g \rightarrow c\bar{c}}}{d y_c d y_{\bar{c}} d p_{c\perp}^2}
\frac{d^2 \Sigma_{\gamma \gamma \rightarrow l\bar{l}}}{d Y d M}
\label{Eq:pocket}
\end{align}
where the photon $\xi_{1,2,3}$ and gluon $x$ momentum fractions read 
\begin{align}
\xi_{1,2} & = \frac{M}{\sqrt{s}} e^{\pm Y} \,, \\
\xi_3,x & = \frac{\sqrt{m_c^2 + p_{c\perp}^2}}{\sqrt{s}} \left( e^{\pm y_c} + 
e^{\pm  y_{\bar c}} \right) \,,
\end{align}
and the effective cross section is given by
\begin{align}
	\sigma^{AA}_\text{eff}(\xi_1, \xi_2, \xi_3)^{-1} = \int d^2 \vec{b} \, 
	T_{\gamma_1 \gamma_2} (\xi_1, \xi_2, \vec{b}) \,  
	T_{g\gamma_3} (\xi_3, \vec{b}) \Theta(b - 2R_A) \,.
	\label{sigma_eff}
\end{align}
In this definition, we have used the overlap function of two photons 
\begin{eqnarray}
T_{\gamma_1 \gamma_2} (\xi_1, \xi_2, \vec{b}) = \frac{1}{\overline{N}_{\gamma_1}(\xi_1) \, \overline{N}_{\gamma_2}(\xi_2)} \, \int d^2 \vec{b}_{\gamma_1} \Theta(b_{\gamma_1} - R_A) \, 
 N_{\gamma_1}(\xi_1,\vec{b}_{\gamma_1}) N_{\gamma_2}(\xi_2, \vec{b}_{\gamma_1} - \vec{b})
\end{eqnarray}
and of a photon and a gluon,
\begin{eqnarray}
T_{\gamma_3 g} (\xi_3, \vec{b}) = \frac{1}{\overline{N}_{\gamma_3}(\xi_3)} 
\int d^2 \vec{b}_p \, \int d^2 \vec{b}_{\gamma_3} \Theta(b_{\gamma_3} - R_A) \, 
N_{\gamma_3}(\xi_3,\vec{b}_{\gamma_3}) \rho_{\rm 2D}(\vec b_p) f_g(\vec{b}_{\gamma_3} - \vec{b_p} - \vec b ) \, .
\end{eqnarray}
The corresponding auxiliary cross sections entering Eq.~(\ref{Eq:pocket}) can be found 
in terms of the parton-level differential SPS cross sections as
\begin{align}
\frac{d^2 \Sigma_{\gamma \gamma \rightarrow l\bar{l}}}{d Y d M^2 } 
 = \xi_1 \overline{N}_{\gamma_1} (\xi_1) \xi_2 \overline{N}_{\gamma_2} (\xi_2) 
\int \frac{d p_{l\perp}^2}{M^2} \frac{d \hat{\sigma}_{\gamma \gamma \rightarrow l\bar{l}}}{d p_{l\perp}^2} \,,
\end{align}
and
\begin{eqnarray}
\frac{d^3 \Sigma_{\gamma g \rightarrow c\bar{c}}}{d y_c d y_{\bar{c}} d p_{c\perp}^2} 
& = & \xi_3 \overline{N}_{\gamma_3}(\xi_3) A x g(x) 
\frac{d \hat{\sigma}_{\gamma g \rightarrow c\bar{c}}}{d p_{c\perp}^2} \,.
\end{eqnarray}


\begin{figure}[tb]
\centering
\includegraphics[width=.49\textwidth]{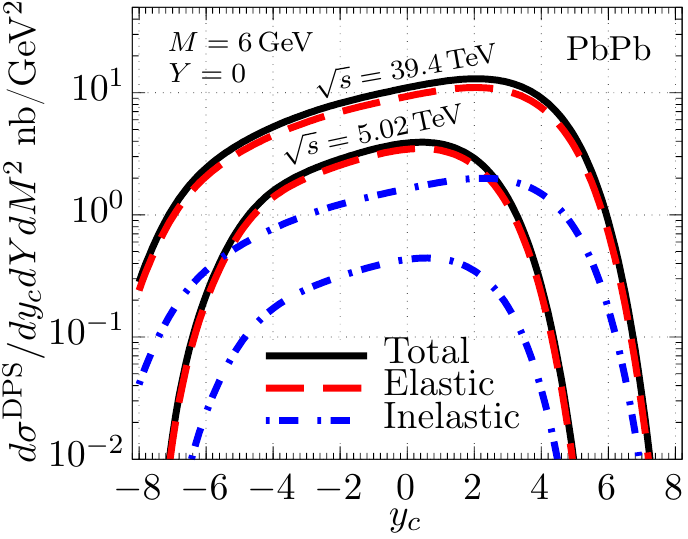} \hfill
\includegraphics[width=.49\textwidth]{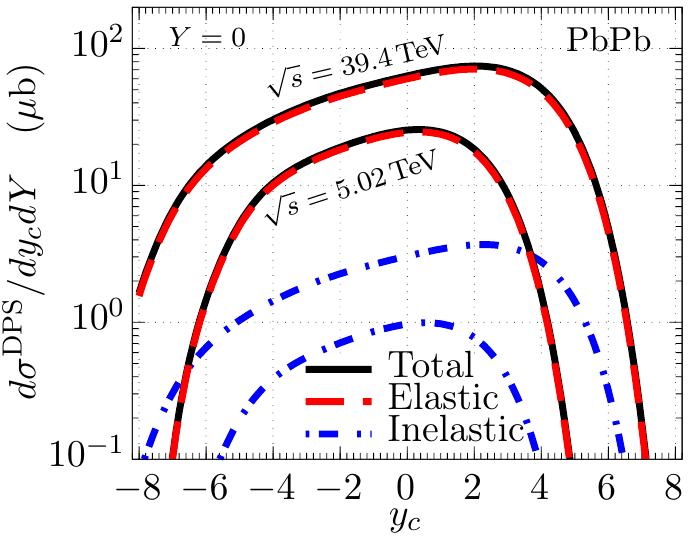}
\caption{
The DPS $c\bar c \mu^+ \mu^-$ production cross section in PbPb UPCs at typical LHC ($\sqrt{s}=5.02$ TeV) and FCC ($\sqrt{s}=39.4$ TeV) energies as a function of $c$-quark rapidity $y_{c}$ with antiquark rapidity $y_{\bar{c}}$ integrated. Both plots have the dilepton rapidity $Y$ fixed, but the left one has the invariant mass $M$ also fixed while the right one integrates on it. Both the elastic and inelastic contributions are included.}
\label{fig:dist_DPS_AA}
\end{figure}

The effective cross section $\sigma^{\rm PbPb}_{\rm eff}$ for $3\gamma + g$ in the initial state in lead-lead UPCs proposed above in Eq.~(\ref{sigma_eff}) is illustrated in Fig.~\ref{fig:sigeff} as a function of the third photon fraction $\xi_3$ for a symmetric configuration with $\xi_1=\xi_2$ as a representative example. This plot is dominated by the distribution of the photon that interacts with the gluon and we will get the smaller effective cross section (larger DPS cross section) when this photon and the other projectile photon are found more easily inside the nucleus. We notice that for small $\xi$ values 
the photons and the gluon rarely overlap and the effective cross section increases, whereas 
for large $\xi$ the photons are usually accumulated in a shell at the projectile nucleus periphery and have a lower probability of interacting with the gluon of the other nuclei in a UPC. A similar effect has been observed in the effective DPS cross section for $2\gamma + 2g$ in the initial state in Ref.~\cite{Huayra:2019iun}, except here the target photon can spread out in a large transverse region than the (second) target gluon there. It means that the effective cross section is flatter at small $\xi_3$. For the same reason the effective cross section is steeper at large $\xi_3$, i.e., the relevance of near miss collisions are diminished as our $\gamma_2$ photon can travel farther. 

Now, we turn to the results of the DPS cross section of $c\bar c l\bar l$
production in $AA$ UPCs for lead-lead nuclei UPCs at $\sqrt{s}=5.02$ TeV, with $A = 208$. As was elaborated in earlier works of Refs.~\cite{Huayra:2019iun,Huayra:2020iib}, 
the DPS cross section cannot be obtained by a simple rescaling of the corresponding 
SPS cross sections as one might naively think which is a result of a nontrivial kinematic
dependence of the effective cross section. 

In Fig.~\ref{fig:dist_DPS_AA} (left panel), we present 
the DPS $c\bar c \mu\bar \mu$ production cross sections differential in $c$-quark rapidity 
(with $y_{\bar{c}}$ being integrated out) at fixed $Y=0$ rapidity and $M=6$ GeV invariant mass 
of the dimuon at two distinct energies, $\sqrt{s}=5.02$ TeV at the LHC and 39.4 TeV, relevant for 
the planned measurements at the Future Circular Collider (FCC). The result of invariant mass 
integration is shown in Fig.~\ref{fig:dist_DPS_AA} (right panel). 
These observables are shown for both elastic (dashed line) and 
inelastic (dash-dotted line) production channels, with the latter being a subleading effect 
as was expected above. 
We have observed a larger DPS $\mu \bar{\mu} c \bar{c}$ production in comparison the DPS $c\bar{c} b \bar{b}$ one found in Ref.~\cite{Huayra:2020iib}, by at least an order of magnitude. This means that, with current LHC integrated luminosity, its possible that such DPS has already happened. The asymmetry, stronger than that of the SPS, is the result of the fact that the effective cross section grows faster for increasing $\xi_3$ at large $\xi_3$ than it grows at small $\xi_3$ for decreasing $\xi_3$.

\begin{figure}[tb]
\centering
\includegraphics[width=.6\textwidth]{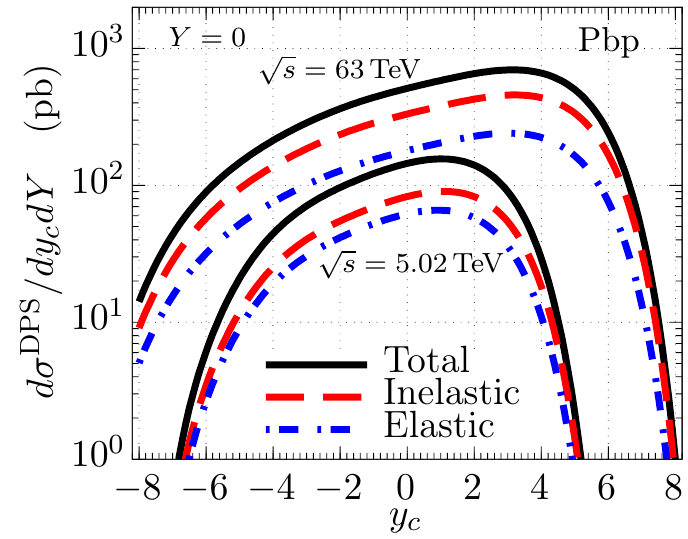}
\caption{The DPS $c\bar c \mu^+ \mu^-$ production cross section in Pbp UPCs at typical LHC ($\sqrt{s}=5.02$ TeV) and FCC ($\sqrt{s}=63$ TeV) energies as a function of $c$-quark rapidity $y_{c}$ with antiquark rapidity $y_{\bar{c}}$ integrated. The plot has the dilepton rapidity $Y$ fixed and invariant mass $M$ integrated. Both the elastic and inelastic contributions are shown.}
\label{fig:dist_DPS_proton}
\end{figure}

For comparison, in Fig.~\ref{fig:dist_DPS_proton} we show the DPS cross section for $c\bar c \mu\bar \mu$ production in lead-proton UPCs differential charm quark rapidity and integrated in anti-quark rapidity, at fixed dimuon $Y = 0$ and integrated invariant mass $M$. We notice that the elastic and inelastic contribution of the DPS cross section in PbPb UPCs is by far larger than Pbp UPCs, in the same channel and kinematics. This demonstrates that the measurement of $c\bar c \mu\bar \mu$ production in PbPb UPCs is advantageous for DPS studies in a mixed $3\gamma+g$ channel at the LHC and future colliders.

\section{Conclusion}
\label{Sect:concl}

In this work, we propose a new way to explore a mixed photon-gluon DPS in $AA$ UPCs which differs from more conventional channels corresponding either pure photon or double-gluon DPS. For this purpose, we estimate the $c\bar c l\bar l$ production  observable differential in charm quark rapidity and dilepton invariant mass and rapidity for the first time in the DPS $3\gamma + g$ channel. Such a process enables one, in particular, to probe the poorly known photon density inside the nucleus and its correlations with the gluon density. For this case, we derived an analogue  of the pocket formula and studied the kinematical dependence of the corresponding double-parton effective cross section.

In this first simplified analysis we do not imply any model for photon-gluon correlations; instead we assume that there are no such correlations. This is a result from the fact that photon have a Weizsäcker--Willians distribution, while the gluon are the ones from standard parton distributions. Therefore, their densities are factorised. This is a different picture from what was done earlier, in the case of $2\gamma + 2g$ initial state (see Ref.~\cite{Huayra:2020iib}), as in there the gluons could be correlated from the fact that they could come from the same nucleon. Our current work can be viewed as a motivation for future measurements of the $c\bar c l\bar l$ final states in $AA$ UPCs given the sizeable cross section we predict at the LHC and FCC. If the future data show a discrepancy with our simplified approach, this will mean that the possible photon-gluon correlations currently neglected in the literature will need to be reevaluated, including paying more attention to the photon as a parton. 

\section*{Acknowledgments}

This work was supported by Fapesc, INCT-FNA (464898/2014-5), and CNPq (Brazil) for EH, EGdO, and BOS. This study was financed in part by the Coordenação de Aperfeiçoamento de Pessoal de Nível Superior -- Brasil (CAPES) -- Finance Code 001. The work has been performed in the framework of COST Action CA15213 ``Theory of hot matter and relativistic heavy-ion collisions'' (THOR). R.P.~is supported in part by the Swedish Research Council grant, contract number 2016-05996, as well as by the European Research Council (ERC) under the European Union's Horizon 2020 research and innovation programme (grant agreement No 668679). 














\begin{thebibliography}{10}

    \bibitem{Manohar:2012pe}
    A.~V.~Manohar and W.~J.~Waalewijn,
    Phys. Lett. B \textbf{713}, 196-201 (2012)
    [arXiv:1202.5034 [hep-ph]].
    
    \bibitem{Diehl:2011yj}
    M. Diehl, D. Ostermeier, and A. Schafer,
    JHEP \textbf{03}, 089 (2012)
    [arXiv:1111.0910 [hep-ph]].
		
	\bibitem{Bartalini:2011jp}
	P.~Bartalini \textit{et al.},
	[arXiv:1111.0469 [hep-ph]].
	
	\bibitem{Diehl:2017kgu}
	M.~Diehl, J.~R.~Gaunt, and K.~Sch\"onwald,
	JHEP \textbf{06}, 083 (2017)
	[arXiv:1702.06486 [hep-ph]].
	
	\bibitem{Bansal:2014paa}
	Bansal, S. \textit{et al.},
	[arXiv:1410.6664 [hep-ph]].
    
    
    
    
    
    
    
    
	
	\bibitem{Aaij:2020smi}
	R.~Aaij \textit{et al.} [LHCb],
	Phys. Rev. Lett. \textbf{125}, no.21, 212001 (2020)
	[arXiv:2007.06945 [hep-ex]].
	
	\bibitem{Aaboud:2018tiq}
	M.~Aaboud \textit{et al.} [ATLAS],
	Phys. Lett. \textbf{790}, 595 (2019)
	[arXiv:1811.11094 [hep-ex]].

	\bibitem{Sirunyan:2017hlu}
	A.~M.~Sirunyan \textit{et al.} [CMS],
	JHEP \textbf{02}, 032 (2018)
	[arXiv:1712.02280 [hep-ex]].
	
	\bibitem{Huayra:2019iun}
	E.~Huayra, E.~G.~de Oliveira and R.~Pasechnik,
	Eur. Phys. J. C \textbf{79}, no.10, 880 (2019)
	[arXiv:1905.03294 [hep-ph]].
		
    \bibitem{vonWeizsacker:1934nji}
    C.~F.~von Weizsacker,
    Z. Phys. \textbf{88}, 612-625 (1934)

    \bibitem{Williams:1934ad}
    E.~J.~Williams,
    Phys. Rev. \textbf{45}, 729-730 (1934)

    \bibitem{Williams:1935dka}
    E.~J.~Williams,
    Kong. Dan. Vid. Sel. Mat. Fys. Med. \textbf{13N4}, no.4, 1-50 (1935)

    \bibitem{Huayra:2020iib}
    E.~Huayra, E.~G.~de Oliveira and R.~Pasechnik,
    Eur. Phys. J. C \textbf{80}, no.8, 772 (2020)
    [arXiv:2003.06412 [hep-ph]].
		
    \bibitem{Rinaldi:2021vbj}
    M.~Rinaldi and F.~A.~Ceccopieri,
    [arXiv:2103.13480 [hep-ph]].

	\bibitem{KlusekGawenda:2010kx}
	M.~Klusek-Gawenda and A.~Szczurek,
	Phys. Rev. C \textbf{82}, 014904 (2010)
	[arXiv:1004.5521 [nucl-th]].

    \bibitem{Klusek-Gawenda:2020eja}
    M.~K\l{}usek-Gawenda, W.~Sch\"afer and A.~Szczurek,
    Phys. Lett. B \textbf{814}, 136114 (2021)
    [arXiv:2012.11973 [hep-ph]].

    \bibitem{Harland-Lang:2021ysd}
    L.~A.~Harland-Lang, V.~A.~Khoze and M.~G.~Ryskin,
    [arXiv:2104.13392 [hep-ph]].

    \bibitem{Brandenburg:2021lnj}
    J.~D.~Brandenburg, W.~Zha and Z.~Xu,
    [arXiv:2103.16623 [hep-ph]].
    
	\bibitem{Vogt:2007zz}
	VOGT, R. \textbf{Ultrarelativistic heavy-ion collisions}. 
	Lawrence Berkeley Laboratory, USA: Elsevier, 2007.


	\bibitem{Jackson:1998nia}
	JACKSON, John David, \textbf{Classical electrodynamics}; 2nd ed. New York, NY: Wiley, 1975.		
	https://cds.cern.ch/record/100964


	\bibitem{Bromley:1967ixa}
	D.~A.~Bromley and J.~Weneser,
	Comments Nucl. Part. Phys. \textbf{1}, no.5, 174-179 (1967)
	
	\bibitem{Mariola:Thesis}
	M.~Klusek-Gawenda \textbf{Production of pairs of mesons, leptons and quarks in ultraperipheral ultrarelativistic heavy ion collisions}. 
	IFJ Polish Academy of Sciences, Poland: 2014.
	 http://rifj.ifj.edu.pl/handle/item/60 
		
	\bibitem{Harland-Lang:2019pla}
	L.~A.~Harland-Lang, A.~D.~Martin, R.~Nathvani and R.~S.~Thorne,
	Eur. Phys. J. C \textbf{79}, no.10, 811 (2019)
	[arXiv:1907.02750 [hep-ph]].
	
	\bibitem{Trzebinski:2019tmk}
	M.~Trzebinski [ATLAS],
	[arXiv:1909.10827 [physics.ins-det]].
	
	\bibitem{Guzey:2014axa}
	V.~Guzey and M.~Zhalov,
	[arXiv:1405.7529 [hep-ph]].
	
	\bibitem{daSilveira:2021bzs}
	G.~G.~da Silveira, V.~P.~Goncalves and G.~G.~V.~Veronez,
	[arXiv:2103.00992 [hep-ph]].
	
    \bibitem{Drees:1988pp}
    M.~Drees and D.~Zeppenfeld,
    Phys. Rev. D \textbf{39}, 2536 (1989)
	
	\bibitem{Eskola:2016oht}
	Eskola, Kari J. and Paakkinen, Petja and Paukkunen, Hannu and Salgado, Carlos A.,
	[10.1140/epjc/s10052-017-4725-9].

	\bibitem{Frankfurt:2010ea}
	L.~Frankfurt, M.~Strikman and C.~Weiss,
	Phys. Rev. D \textbf{83}, 054012 (2011)
	[arXiv:1009.2559 [hep-ph]].

\end{thebibliography}
\end{document}